\documentstyle[preprint,aps]{revtex}
\begin{document}
\tighten
\title{ Interaction of $\eta$-meson  with light nuclei}
\author{ V. B. Belyaev}
\address{Joint Institute  for Nuclear Research, Dubna, 141980, Russia}
\author{ S. A. Rakityansky\footnote{Permanent address:
Joint Institute  for Nuclear Research, Dubna, 141980, Russia},
 S. A. Sofianos, and M. Braun}
\address{ Physics Department, University of South Africa,
P.O.Box 392, Pretoria 0001, South Africa}
\author{W. Sandhas}
\address{Physikalisches Institut, Universit\H{a}t
 Bonn, D-53115 Bonn, Germany}
\date{\today}
\maketitle
\begin{abstract}
A microscopic treatment of $\eta$--nucleus scattering is presented.
When applying the underlying exact integral equations, the excitation
of the target are neglected, and an input $\eta N$ amplitude is chosen
which  reproduces the $S_{11}(1535)$ resonance. It is shown that
the $\eta$--nucleus scattering lengths are quite sensitive to the
$\eta N$ parameters and the nuclear wave functions. For a special
choice of the parameters an $\eta {^4\!H}e$ quasi--bound state occurs.
\end{abstract}
\section{Introduction}
There are at least three main reasons which motivate studies concerning
$\eta$--nucleus systems.\\

The first one is of a fundamental character and is related to the
possibility of studying the quark structure of  $\eta$--mesons
 and the nucleon $S_{11}(1535)$--resonance,  and the role played by
the strange quarks in these systems.
Recently, in experiments with $N\overline N$ annihilation into $\phi$ and
$\omega$  channels \cite{Sapo}, it was established that there is a strong
polarization of the strange sea--quarks in nucleons. It is, therefore,
interesting to investigate possible  manifestation of this phenomenon also in
$\eta$--nucleus systems.\\

The second reason concerns a pure nuclear problem,  namely,  the possible
formation of $\eta$~--~nucleus bound states. It was argued by
Haider and Liu \cite{Haid} that, due to the rather strong  attraction
in the $\eta N$--interaction at low energies,  the $\eta$--meson,
when immersed in the nucleus, might create a system
which in some respects is analogous to a hypernucleus. Estimations
in  \cite{Haid} indicated that such a possibiltity
exists for  nuclei with atomic number $A\ge12$.\\

Thirdly,  Charge Symmetry Breaking (CSB) effects can be studied in
processes involving $\eta$--production. Indeed, CSB was experimentally
 observed at Saturne \cite{Fras} in the reaction
\begin{equation}
     d + d\longrightarrow {^4\!H}e +\pi^0,
\label{dda}
\end{equation}
near the threshold of  $\eta$--production. It was found that the
production of pions was much greater than the CSB   expected from
electromagnetic interaction. The experimental cross--section is equal
to 0.97 $pb/sr$ (at $E_d=1100$ MeV), while the one obtained from
electromagnetic considerations is 0.003 $pb/sr$ \cite{Cheu}, the latter
estimate
being made at 600 MeV. \\

A possible explanation of CSB in the above reaction is that it can be
attributed to the $\eta-\pi^0$ mixing, i.e., the process (\ref{dda}) can
proceed via  intermediate production of a pure  $\eta^0$ meson state  with
isospin $I=0$,
\begin{equation}
     d + d\longrightarrow {^4\!H}e +\eta^0,
\label{ddb}
\end{equation}
which is an allowed reaction. The state $|\eta^0\rangle$ can be expressed
as a linear combination of the physical ones, $|\eta\rangle$ and
$|\pi^0\rangle$,
\begin{equation}
    |\eta^0\rangle={\displaystyle \frac{1}{\sqrt{1+\lambda^2}}
     (\,|\eta\rangle + \lambda |\pi^0\rangle\,) },
\label{etapi}
\end{equation}
where the parameter $\lambda$ characterises the $\eta\,-\, \pi^0$  mixing.
{}From Eq. (\ref{etapi}) it follows that $\lambda$ is the
proportionality constant between the amplitudes \cite{W}
\begin{equation}
   f(dd\rightarrow\alpha\pi^0) \sim\lambda f(dd\rightarrow\alpha\eta).
\label{fdd}
\end{equation}
The process (\ref{dda}), therefore, is no longer forbidden
and one can expect a cross section  much larger than the
electromagnetic one. \\
Based on the above ideas, the authors of Ref. \cite{Coon}  obtained
0.12 $pb/sr$ for the cross--section of the process (\ref{dda}) as compared
to the experimental value of 0.97 $pb/sr$. To explain this discrepancy,
fully microscopic few--body calculations, which take into account the
$\eta-\pi^0$ mixing, should be performed.\\

We would also like to mention the possibility of studying processes of the
type
\begin{equation}
     d + d\longrightarrow {^4\!H}e^*\,({\cal J}^P,0) +\pi^0 .
\label{dd4}
\end{equation}
Here, ${^4\!H}e^*\,({\cal J}^P,0)$ denotes excited states of ${^4\!H}e$
nucleus with isotopic spin $I=0$ and different angular momenta
and parities.  These reactions might provide us with large cross sections
due to the extended size of the  excicted ${^4\!H}e$ nucleus.\\

Having the above in mind, and in order to get some insight into the
physics of  $\eta$--nucleus systems,  we undertook microscopic
investigations concerning the low--energy behaviour of the
$\eta$--nucleus elastic scattering amplitude  and  the positions of
poles of the corresponding t-matrices in the complex $k$--plane. Another
issue addressed is up to  what extent the  attraction of the two--body
$\eta N$ interaction should be enhanced in order  to produce quasi-bound
states  in the $\eta$--nucleus system.
\section{The Method}
Our method is based on the so--called Finite--Rank--Approximation (FRA) of
the Hamiltonian proposed as an alternative to the multiple scattering
theory of pion-nucleus interaction~\cite{Bely}. We, therefore, start
by outlining this approach.\\

Consider the  scattering of an $\eta$-meson from a nucleus of atomic
number A. The total Hamiltonian of this system is
\begin{equation}
	  H = H_0 + V + H_A ,
\label{H}
\end{equation}
where $H_0$ is the kinetic energy operator (free Hamiltonian) of the
$\eta$--nucleus motion , $ V = V_1 + V_2 +\cdots+ V_A $   is the sum of
$\eta N$--potentials, and $H_A$  is the total Hamiltonian of the nucleus.
Introducing the Green function
\begin{equation}
           G_A(z)=\frac{1}{z-H_0-H_A},
\label{GA}
\end{equation}
we obtain the following equation for the transition operator
\begin{equation}
          T(z)=V+V G_A(z)T(z).
\label{T}
\end{equation}
For further developments, we introduce the (auxiliary) transition
operator $T^0(z)$ via
\begin{equation}
          T^0(z)=V+V G_0(z)T^0(z),
\label{T0}
\end{equation}
where
\begin{equation}
          G_0(z)= \frac{1}{z-H_0}
\label{G0}
\end{equation}
is the free Green function. From the definition of $G_0(z)$ and $V$,
it is clear that $T^0$ describes the scattering of mesons by a
nucleus in which the position of the nucleons is fixed. The operator $T^0$
differs from the usual fixed center t-matrix in two
respects. Firstly,  Eq. (\ref{T0}) contains the kinetic energy operator
$H_0$ describing the motion of the $\eta$--meson with respect to the
center of mass  of the target, secondly, the energy argument
of $T^0$ is taken to be that of the total energy $z$ of the system.\\

Using the resolvent equation
\begin{equation}
      G_A(z)=G_0(z)+G_0(z)H_A G_A(z),
\label{GAA}
\end{equation}
we easily infer from  (\ref{T})  and (\ref{T0})
\begin{equation}
             T(z) =T^0(z)+T^0(z)G_0(z)H_AG_A(z)T(z),
\label{TT0}
\end{equation}
a form suitable for the low--energy approximation on which our approach is
essentially based. The spectral decomposition of the nuclear Hamiltonian
$H_A$ reads
\begin{equation}
      H_A=\sum_n {\cal E}^A_n |\Psi^A_n\rangle\langle \Psi^A_n|
         +\int E|\Psi^A_E\rangle\langle \Psi^A_E|\,dE\,,
\label{HE}
\end{equation}
where $|\Psi^A_n\rangle $ are the bound--state eigenfuntions of  $H_A$
with ${\cal E}_n$ being the correponding energies. Since we are
interested in processes at energies far below the excited states or
break--up thresholds of the nuclei, we can neglect all contributions
to (\ref{HE}) except the ground--state part. That is, we use the
approximation
\begin{equation}
     H_A\approx {\cal E}_0^A|\Psi_0^A><\Psi_0^A| \label{Happrox}.
\label{HAE}
\end{equation}
Inserting Eq. (\ref{HAE}) into Eq. (\ref{TT0})  and sandwiching with
$|\Psi_0^A\rangle$, we obtain the Lippmann--Schwinger--type equation
\begin{equation}
   T(\vec k',\vec k;z) =  T^0(\vec k',\vec k;z)
    + {\cal E}_0^A\int \frac{d^3 q}{(2\pi)^3}
	\frac{T^0(\vec k',\vec q;z)}{(z-{\displaystyle\frac{q^2}{2\mu}})
     (z-{\cal E}_0-{\displaystyle \frac{q^2}{2\mu})}}\,T(\vec q,\vec k;z),
\label{T1}
\end{equation}
where
\begin{equation}
     T^0(\vec k',\vec k;z)= \int \,d^{3(A-1)}r |\Psi^A_0(\vec r)|^2
           T^0(\vec k',\vec k;\vec r;z),
 \label{T0int}
\end{equation}
Here $\vec r$ represents all nuclear Jacobi coordinates.\\

{}From a practical point of view it is convenient to rewrite Eq.
(\ref{T0}) by using the Faddeev-type decomposition
$$
T^0(z)= \sum^A_{i=1} T^0_i(z).
$$
Introducing the operators
$$
t_i(z)=V_i+V_i\frac{1}{z-H_0}t_i(z),
$$
we finally  get for the Faddeev components $T^0_i$  the following system
of integral equations
\begin{eqnarray}
\nonumber
 T^0_i (\vec{k}',\vec{k};\vec{r};z)
&=& t_i(\vec{k}',\vec{k};\vec{r};z) \\
  &+& \int \frac {d^3k''}{(2\pi)^3}
\frac {t_i(\vec{k}',\vec{k}'';\vec{r};z)} {z -{\displaystyle \frac
{k''^2}{2\mu}}} \sum_{j\neq i} T^0_j(\vec{k}'',\vec{k};\vec{r};z) .
\label{t0i}
\end{eqnarray}

\noindent
The  amplitude $t_i$ describes the scattering of the $\eta$-meson off
 the $i$-th  nucleon. It is expressed in terms of the corresponding
two-body $t_{\eta N}$-matrix via
$$
t_i(\vec k',\vec k;\vec r;z)=t_{\eta N}(\vec k',\vec k;z)
 \exp\left[{{\displaystyle i(\vec k-\vec k')\cdot\vec r_i}}\right],
$$
where $\vec r_i$ is the  vector from the nuclear center of mass to the
 $i$-th nucleon and can be expressed in terms of the Jacobi coordinates
$\{\vec r\}$.\\

To get the nuclear wave function we used the so-called
Integro--Differential Equation--Approach (IDEA)\cite{IDEA1,IDEA2}.
In this approach the wave-function of a nucleus consisting of A nucleons
is decomposed in  Faddeev-type components
$$
     \Psi ({\bf r})=\sum_{i<j\leq A} \,\psi_{ij}({\bf r}),
$$
obeying
$$
    (h_0-E)\psi_{ij}({\bf r})=-V(r_{ij})\,\sum_{k<l\leq A}
   \,\psi_{kl}({\bf r}).
$$
Here $h_0$ is the kinetic energy operator of the nucleus, and $V_{ij}$ is
the potential between nucleons $i$ and $j$. These components are written as
a product of a harmonic polynomial $H_{[L_m]}$ and an unknown function $P$
$$
          \psi_{ij}({\bf r})=
          H_{[L_m]}({\bf r}) \frac{1} {\rho^{{\cal L}_0+1}} P(z,\rho),
$$
where $P$ depends only on the pair separation $r_{ij}=\sqrt{\rho(1+z)/2}$
and the hyperradius $\rho=\left[ 2/A\sum r^2_{ij}\right ]^{1/2}$.
Eventually, one has to solve an integro-differential equation for this
function $P$ which, for an A-boson system, reads \cite{IDEA1,IDEA2}
\begin{eqnarray}
 \Biggl \{ \frac{\hbar^2}{m} \bigg [
    &-& \frac{\partial^2}{\partial \rho^2}
    +\frac{ {\cal L}_0({\cal L}_0+1)}{\rho^2}-\frac{4}{\rho^2}
  \frac{1}{W_0(z)}\frac{\partial}{\partial z}(1-z^2)W_0(z)
  \frac{\partial}{\partial z}\bigg ] \nonumber \\
&+& \frac{A(A-1)}{2}V_0(\rho) -E \Biggr\} P(z,\rho)=\nonumber \\
&-&\left[ V\left(\rho\sqrt{\frac{1+z}{2}}\right)-V_0(\rho)\right]
   \left\{P(z,\rho)+\int_{-1}^{+1}f_0(z,z')P(z',\rho)dz'\right\}.\nonumber
\label{ID}
\end{eqnarray}
$V_0(\rho)$ is the  hypercentral potential. The details of the
method which takes into account  the two-body correlations  exactly
can be found  in Refs. \cite{IDEA1,IDEA2}.
\section{Results}
To  solve Eq. (\ref{T1})  we need,  as an input to Eq. (\ref{t0i}), the
two--body
$\eta N$--amplitude (or the corrresponding t-matrix) and the ground state
wave function $\Psi_0^A$  of the nucleus.  For $t_{\eta N}$ we employ
the  $S$-wave separable form
\begin{equation}
     t_{\eta N}(k,k',z)=\frac{\lambda}{(k^2+\alpha^2)
         (z-E_0+i\Gamma/2)(k'^2+\alpha^2)}
\end{equation}
where $E_0$ and $\Gamma$ are the position and width of the $S_{11}$
resonance. Different sets of parameters $\alpha$ and $\lambda$
are chosen such that the scattering lengths in \cite{Bhal} are
reproduced.

For such a form, the equation for $T^0$ can be solved analytically.
The results  thus  obtained are given in Table 1. The required nuclear
wave functions were constructed using the Integro--differential Equation
with the Malfliet-Tjon I+III (MT) nucleon--nucleon potential \cite{MT}.
As can be seen there is a strong dependence of the resulting $\eta$--nucleus
scattering length on the two-body input. A large $\eta N$--scattering length
generates a strong attraction of the $\eta$--meson by the three-- and
four--nucleon  systems. \\

In Table 2 we present $\eta$--nucleus scattering lengths obtained by
using two different sets of the nuclear wave functions, namely, the one
constructed using the IDEA with MT potential and a phenomenological
one of  Gaussian form chosen to  reproduce the root mean square radii
of the nuclei obtained via the IDEA. We see that the considerable
difference found for $^2H$ practically vanishes when going over to
$^4\!He$.\\

\begin{table}
\caption[]{ The $\eta$-nucleus scattering lengths (in fm)
for  9 combinations of the range parameter
$\alpha$ and the $a_{\eta N}$  scattering lenght \cite{Bhal,Wilk,Benn}.}
\begin{tabular}{|c|c|c|c|c|}
\hline
\hline
\multicolumn{1}{|l|}{}  &
\multicolumn{1}{l|} { $ \alpha$=2.357 (fm$^{-1}$)} &
\multicolumn{1}{l|} { $\alpha$ =3.316 (fm$^{-1}$)} &
\multicolumn{1}{l|} { $\alpha$ =7.617 (fm$^{-1}$)} &
\multicolumn{1}{c|} { $\alpha_{\eta N}$ (fm)}
\\ \hline
 $ ^2H$   & 0.65+$i$0.86 & 0.62+$i$0.91  &  0.54+$i$0.97   &             \\
 $ ^3H$   & 0.81+$i$1.90 & 0.66+$i$1.98  &  0.41+$i$2.00  &0.27+$i$0.22 \\
$^3He$    & 0.79+$i$1.90 & 0.64+$i$1.98 &  0.39+$i$2.00   &            \\
$^4He$    & 0.23+$i$3.54 & -0.40+$i$3.43 & -0.96+$i$2.95  &            \\
\hline
 $ ^2H$   & 0.74+$i$0.77 &0.73+$i$0.83   & 0.67+$i$0.91    &           \\
 $ ^3H$   & 1.12+$i$1.82 &0.99+$i$1.96   & 0.71+$i$2.07    &0.28+$i$0.19\\
 $^3He$   &1.10+$i$1.83 &0.97+$i$1.91   & 0.68+$i$2.07    &           \\
 $^4He$   & 0.96+$i$3.99&0.06+$i$4.13   & -0.91+$i$3.62   &           \\
\hline
 $ ^2H$     & 1.38+$i$2.27&1.05+$i$2.45   & 0.57+$i$2.34    &           \\
 $ ^3H$   & -1.44+$i$4.60& -1.72+$i$3.76  & -1.56+$i$3.00    &0.55+$i$0.30\\
 $^3He$   & -1.42+$i$4.54&-1.68+$i$3.74   & -1.53+$i$2.99    &\\
 $^4He$   & -3.66+$i$1.78 & -2.96+$i$1.39 & -2.42+$i$1.25   &\\
\hline
\end{tabular}
\end{table}
\begin{table}
\caption[]{The $\eta$-nucleus scattering lengths (in
fm) for different nuclear wave functions. In all cases
$a_{\eta N}$=0.55+$i$0.30 fm.}
\begin{tabular}{|c|c|c|}
\hline
 Model & IDEA
  &
  Gaussian \\ \hline\hline
 $ ^2H$   &  1.05+$i$2.45 &   0.57+$i$2.14  \\
 $ ^3H$   & -1.72+$i$3.76 &  -1.42+$i$4.32  \\
$^3He$    & -1.68+$i$3.74 &  -1.39+$i$4.28   \\
$^4He$    & -2.96+$i$1.39 &  -3.07+$i$1.48    \\
\hline
\end{tabular}
\end{table}
To find  at which values of the  $\eta N$--scattering length a quasi--bound
state appears in the $\eta$--nucleus system, we write the
$\eta- N$  scattering length in terms of  two parameters $g_1$ and $g_2$
$$
a_{\eta N}=(0.55 g_1+i0.30 g_2)\,\, {\rm fm}.
$$
Note that  $g_1=g_2=1$ is just one of the choices in Table 1.
These parameters are varied until the $\eta$-nucleus $T$-matrix
exhibits a pole for $Re\, E=0$. Preliminary results of this
search with Gaussian type wave functions are given in Table 3.
\begin{table}
\caption[]{Values of $g_1$  at which a  quasi--bound state appears.}
\begin{tabular}{|c|c|c|c|c|}
\hline
\hline
 &$\alpha= 2.357$ (fm ${}^{-1}$)  &
 $\alpha= 3.316$ (fm ${}^{-1}$)  &
  $\alpha=7.617$ (fm ${}^{-1}$) &
  g${}_2$  \\
\hline
d & 1.610 & 1.613 &   1.547 & 0 \\
\cline{2-5}
& 1.654 & 1.566 & 1.535 & 1 \\
\hline
t & 1.293 & 1.124 & 0.996 & 0 \\
\cline{2-5}
& 1.361 & 1.310 & 1.260 & 1 \\
\hline
${}^3$He & 1.302 & 1.204 & 1.130 & 0 \\
\cline{2-5}& 1.330 & 1.221 & 1.144 & 1 \\
\hline
${}^4$He & $1.075$ & $0.992$ & $0.910$ & $0$ \\
\cline{2-5}
& 0.955 & 0.911 & 0.899 & $1$   \\
\hline
\end{tabular}
\end{table}
It is  seen that an $\eta-\,{^4\!H}e$
quasi--bound state for realistic values of $a_{\eta N}$ can exist. In all
other cases, the physical $\eta N$ input, has to be modified in order
to be able to support a corresponding quasi--bound state.\\

Finally, we  mention that using the auxiliary matrix $T^0$ instead of
the full solution of Eq. (\ref{T1}), we obtained completely different
results.
\section{Conclusions}

We have investigated   $\eta A$ systems with $A\le4$ in the
framework of a few-body microscopic approach by using  different
two-body input parameters and nuclear bound state wave functions.  The
results can be summarized as follows:
\begin{itemize}
\item
 There is a remarkable increase of attraction  with increasing atomic
 number.
\item
 The same tendency  appears for all nuclei when the $\eta N$
scattering length is increased.
\item
It is important in calculating  $\eta A$  observables
to use microscopically constructed wave functions.
\item
The fixed scatterer approximation,
corresponding to the use of $T^0$ instead of $T$, is inadequate in
describing the $\eta$--nucleus system.
\item
 There is a strong indication that the $\eta\, {^4\!H}e$ system
can form a quasi--bound state.
\end{itemize}


\end{document}